\documentclass[a4paper,11pt]{article}
\pdfoutput=1 

\usepackage{jinstpub} 
\usepackage{acronym}
\newcommand{\isotope}[2]{\ensuremath{{^{#1}\textnormal{#2}}}}
\newcommand{\reaction}[4]{\ensuremath{#1(#2,#3)#4}}
\newcommand{\geant}{\ensuremath{\textsc{Geant4}}}

\newcommand{\labr}{LaBr$_{3}$:Ce}

\newcommand{\tref}{Reference~}
\newcommand{\tfig}{Figure~}
\newcommand{\ttab}{Table~}

\title{\boldmath Unfolding of sparse high-energy $\gamma$-ray spectra from LaBr$_{3}$:Ce  detectors}

\author[a]{P.-A.~S\"{o}derstr\"{o}m,\note{Corresponding author.}}
\author[a]{L.~Capponi,}
\author[a]{V.~Iancu,}
\author[a,b,c]{D.~Lattuada,}
\author[a]{A.~Pappalardo,}
\author[a,d]{G.~V.~Turturic\u{a},}
\author[a]{E.~A\c{c}{\i}ks\"{o}z,}
\author[a]{D.~L.~Balabanski,}
\author[a]{P.~Constantin,} 
\author[a,c]{G.~L.~Guardo,}
\author[d,e]{M.~Ilie,}
\author[a,d]{S.~Ilie,}
\author[a]{C.~Matei,} 
\author[a,d]{D.~Nichita,}
\author[a,d]{T.~Petruse,}
\author[a,d]{and A.~Spataru}

\affiliation[a]{Extreme Light Infrastructure-Nuclear Physics (ELI-NP),
Horia Hulubei National Institute for R\&D in Physics and Nuclear Engineering (IFIN-HH),
Reactorului 30, 077125 Bucharest-M\u{a}gurele, Romania}
\affiliation[b]{Universit\'{a} di Enna Kore, Enna 94100, Italy}
\affiliation[c]{Istituto Nazionale di Fisica Nucleare, Laboratori Nazionali del Sud, Catania 95125, Italy}
\affiliation[d]{Politehnica University of Bucharest, Splaiul Independentei 313, 060042 Bucharest, Romania}
\affiliation[e]{Tandem Accelerators Department (DAT), Horia Hulubei National Institute for R\&D in Physics and Nuclear Engineering (IFIN-HH),
Reactorului 30, 077125 Bucharest-M\u{a}gurele, Romania}

\emailAdd{par.anders@eli-np.ro}

\abstract{Here we report on the characterization of one of the large-volume LaBr$_{3}$:Ce detectors for the ELIGANT project at ELI-NP. The main focus of this work is the response function for high-energy $\gamma$ rays of such detectors. In particular, we compare a selection of unfolding methods to resolve small structures in $\gamma$-ray spectra with high-energies. Three methods have been compared using $\gamma$-ray spectra with energies up to 12~MeV obtained in an experiment at the 3~MV Tandetron\texttrademark\ facility at IFIN-HH. The results show that the iterative unfolding approach gives the best qualitative reproduction of the emitted $\gamma$-ray spectrum. Furthermore, the correlation fluctuations in high-energy regime from the iterative method are two orders of magnitude smaller than when using the matrix inversion approach with second derivative regularization. In addition, the iterative method is computationally faster as it does not contain large matrix inversions. The matrix inversion method does, however, give more consistent results over the full energy range and in the low-statistics limit. Our conclusion is that the performance of the iterative approach makes it well suitable for semi-online analysis of experimental data. These results will be important, both for experiments with the ELIGANT setup, and for on-line diagnostics of the energy spread of the $\gamma$-ray beam which is under implementation at ELI-NP.}

\keywords{Gamma detectors, Analysis and statistical methods}

\begin{document}
\maketitle
\flushbottom

\section{Introduction}

The \ac{ELI-NP} \cite{gales_rpp} facility currently under implementation in Romania will be a unique European laboratory for photonuclear physics. One of the projects being constructed under the \ac{ELI-NP} umbrella is  \ac{ELIGANT}. \ac{ELIGANT}, especially the \ac{ELIGANT-GN} \cite{gant_tdr,gant_matt,matt_aip} setup, will focus on competing $\gamma$-ray and neutron emission in photonuclear reactions. The goal of \ac{ELIGANT-GN} is the detailed study of the high-energy photo-excitation response of atomic nuclei with focus on the \ac{GDR}, \ac{PDR}, and similar structures through simultaneous measurements of neutron and $\gamma$-ray decay-channels. The narrow-bandwidth nature of the proposed $\gamma$-ray beam \cite{diagnostics_tdr} will provide a unique opportunity to scan the \ac{GDR} and \ac{PDR} with a well defined energy, in many cases smaller than the typical energy resolution of scintillator detectors. 
This means that the \ac{ELIGANT} collaboration will be able to, for example, study in detail the decay branching of the \ac{GDR} fine structure \cite{PhysRevLett.107.062502} and the \ac{PDR} to the ground, $0^{+}$, state and the first excited, $2^{+}$, state for even-even nuclei. The detection of different $\gamma$ branches to the first $0^{+}$ and $2^{+}$ states is straightforward in some doubly magic cases, like $^{208}$Pb, where the excitation energy of the first excited state is around 4~MeV. However, if we want to explore similar topics in slightly deformed nuclei, the energy difference between the ground state transition and the transition to excited states will be smaller and the signal of interest may be hidden under the response of the $\gamma$-ray detector. 

Another important aspect in the evaluation of future experimental data is the properties of the beam itself. It was, indeed, noted in a report by the \ac{IAEA} \cite{iaea_tecdoc_1178} that one of the major sources of observed systematic disagreements in the evaluation of photonuclear data was the differences in photon spectra where the cross-sections were derived by unfolding of the data. This triggered a new large-scale experimental campaign for re-measuring several key elements \cite{iaea_2019}. At \ac{ELI-NP} a large beam diagnostics program is under development with several instruments being implemented. For example, one proposed instrument that will measure the absolute energy as well as the energy spread of the $\gamma$-ray beam is a large volume \ac{HPGe} or \labr\ detector with anti-Compton shield placed directly in the beam, following an attenuator \cite{diagnostics_tdr}. In order to have control of the beam properties it is important to quickly understand the underlying beam spectrum from the measured spectrum of the monitoring detector. Another proposed instrument based on a similar principle, but using a Compton scattered beam component instead of the attenuated main beam, has been reported in \tref\cite{TURTURICA201927}. In that work a \ac{HPGe} detector was used for monitoring the beam intensity from comparing experimental data to \geant\ simulations.  For both of these types of instruments it can be desirable to have a fast and accurate evaluation for control of beam parameters. 

In previous work, the \ac{ELIGANT-GN} detectors have been thoroughly characterized in the low-energy regime in the context of the \ac{ELIGANT-GG} \cite{eligant-gg} setup for studies of competitive double-$\gamma$ decay \cite{maria_gm_double_gamma_phd,walz_nature}. In the high-energy regime, the response function and linearity of larger volume \labr\ detectors were investigated in the energy range $6-38$~MeV by direct measurements at the NewSUBARU synchrotron radiation facility \cite{labr_linearity}. 

Here we will report on an experiment for testing the high-energy response of the \ac{ELIGANT-GN} detectors with particular focus on different methods of unfolding the experimental spectra to resolve small structures. 

\section{Experiment}

The experiment was performed at the 3~MV Tandetron\texttrademark\ facility at the \ac{IFIN-HH}, M\u{a}gurele, Romania \cite{ifin_3mv}. The $\gamma$ rays used for this study were obtained from a 1.05~MeV proton beam with an average beam current of 11.6~$\mu$A impinging on a composite target of Al and CaF$_{2}$ with a mass ratio of 99\% and 1\%, respectively \cite{turturica_dgamma_paper}. This produced $\gamma$ rays in three energy groups, around 2~MeV and 11~MeV from the \reaction{\isotope{27}{Al}}{\mathrm{p}}{\gamma}{\isotope{28}{Si}} reaction and around 7~MeV from the \reaction{\isotope{19}{F}}{\mathrm{p}}{\gamma\alpha}{\isotope{16}{O}} reaction. 
One \ac{ELIGANT-GN} detector was used for this measurement, which consisted of a $3\times3$~inch \labr\ crystal coupled with a Hamamatsu R11973 \ac{PMT} and a AS20 voltage divider. The signals from the detectors were read out by a CAEN v1730 digitizer using a sampling frequency of 500~MS/s and a resolution of 14 bits. 

\begin{figure}[ht]
 \centering
 \includegraphics[width=0.446\columnwidth]{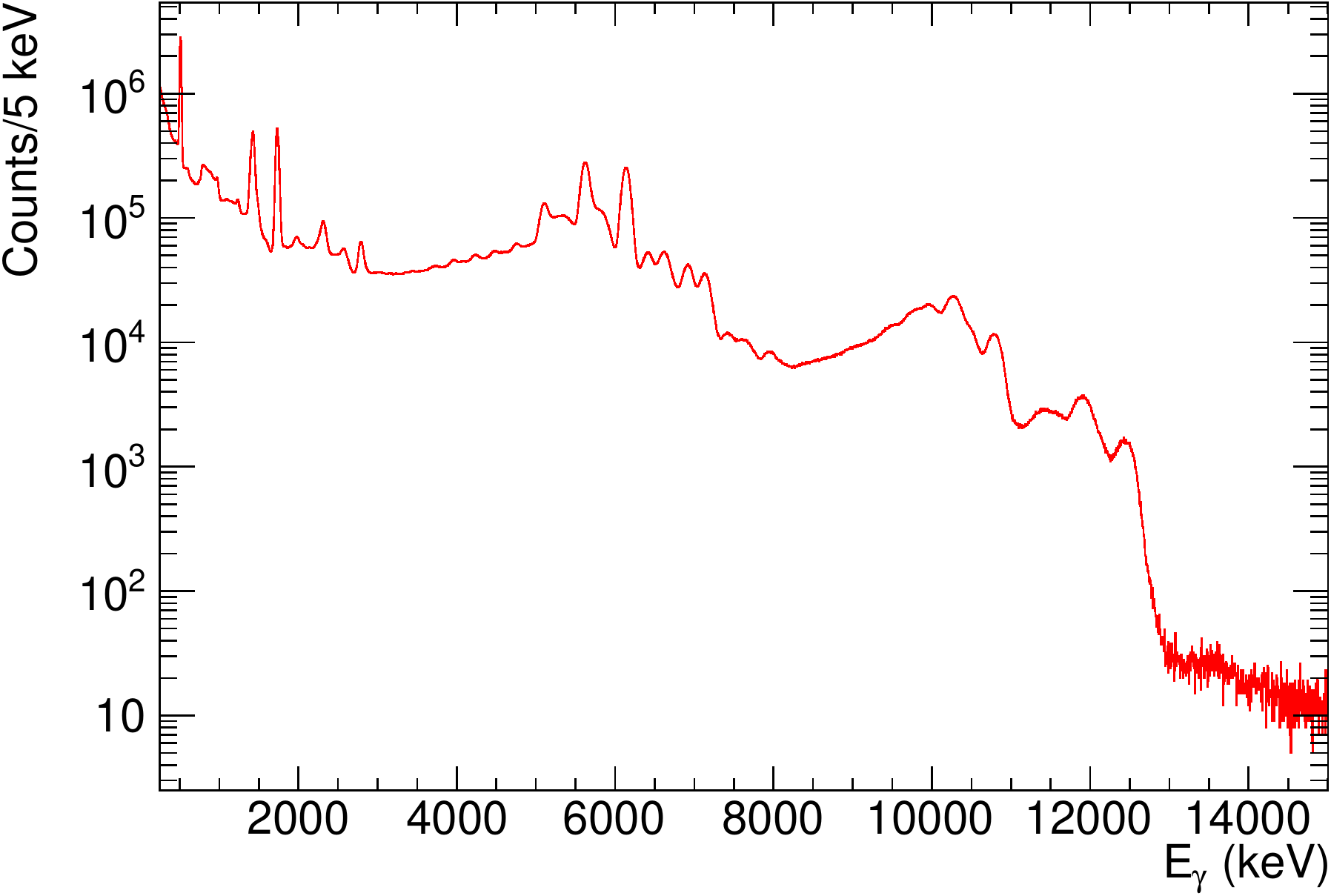}
 \includegraphics[width=0.534\columnwidth]{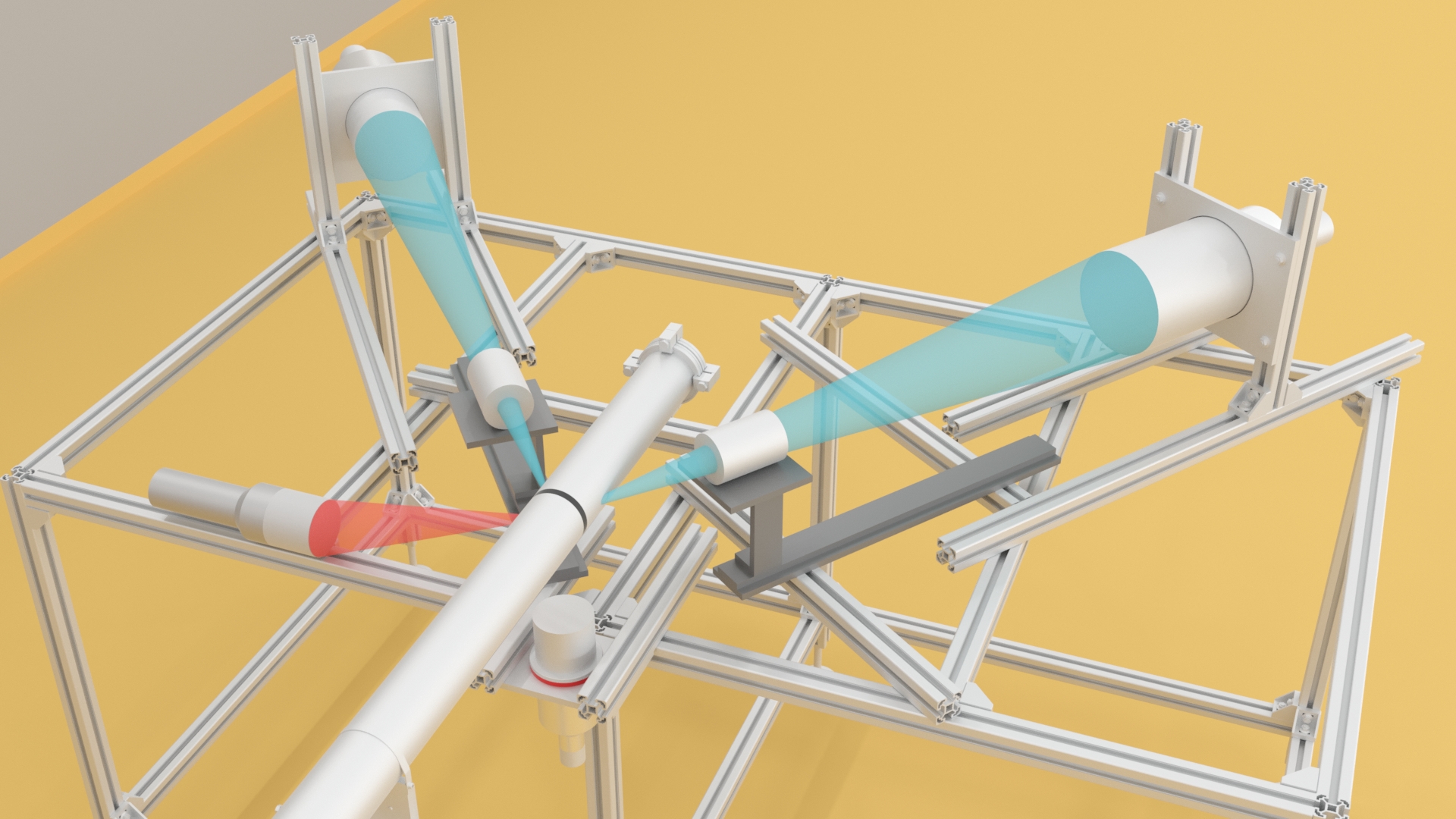}
 \caption{(Left) Energy spectrum obtained in the \labr\ detector in the 250~keV to 15~MeV range. (Right) Experimental setup with the measurements for industrial applications \cite{turturica_dgamma_paper} shown in blue and the measurement for high-energy response of the ELIGANT detectors shown in red.\label{fig:setup}}
\end{figure}

The experiment was performed concurrently to another experiment aiming for effective $Z$ evaluation of an unknown material \cite{turturica_dgamma_paper}. Thus, the geometry of the experimental setup, shown in \tfig\ref{fig:setup}, was such that the \labr\ detector was not aligned with the front-face of the detector in the direction of the reaction target. Instead, the \ac{ELIGANT-GN} detector was placed at an angle of 90~degrees relative to the beam axis at a distance of 40~cm from the target, 20~cm below the beam. This was taken into careful consideration in the \geant\ simulations used to construct the detector response. As a consequence, the single- and double escape peaks in the energy spectrum were approximately a factor of 50\% larger relative to what is expected when the detector is directly facing the source, as in the \ac{ELIGANT-GN} design. The energy spectrum obtained from this experiment is shown in \tfig\ref{fig:setup}.

\section{Detector response}

 As a first step in the calibration of the detector, a linear calibration was carried out only using the 511~keV peak from positron annihilation and the 1779~keV peak from the $2^{+}\to0^{+}$ transition in $^{28}$Si from the \reaction{\isotope{27}{Al}}{\mathrm{p}}{\gamma}{\isotope{28}{Si}} reaction. 
 The full energy calibration was performed using the strongest peaks of the known $\gamma$-ray transitions following the two reactions: 511~keV, 1779~keV, 6130~keV, 6917~keV, 7117~keV, and 10760~keV. In addition, we also used the first and second escape peak of the 6130~keV transition , and the first escape peaks of the 6917~keV, 7117~keV, 10760~keV, and 12331~keV transitions. For this purpose, a second-order polynomial was used to correct the linear calibration.


The basic idea behind the unfolding of experimental data is that if there is a histogram of an emitted spectrum, $\vec{x}$, separated into $n$ bins, this spectrum can be measured by a detector with an $n \times m$ response matrix, $\mathbf{A}$, resulting in a measured spectrum, $\vec{y}$, in a histogram consisting of $m$ bins, as 
\begin{equation}
 \vec{y} = \mathbf{A}\vec{x}.
\end{equation} 
If the detector response is linear, as it can be approximated in the low-energy regime, the interpretation of $\vec{y}$ is straightforward as each bin directly corresponds to an emitted energy given a finite detector resolution. If this data is non-linear, as is the case in the high-energy regime, $\vec{y}$ contains a large contribution from physics processes such as Compton scattering and electron/positron escape in  addition to the full energy deposition. In these cases, the response $\mathbf{A}$ has to be properly characterized to understand the data. For this work, we used \geant\ simulations \cite{2003NIMPA.506..250G} (version 10.05) to obtain $\mathbf{A}$ via the dedicated \ac{GROOT} software developed for \ac{ELI-NP} \cite{groot}. For this purpose a desktop computer with 7.7~GB of memory and an Intel\textregistered\ Xeon\textregistered\ Processor E5-1620 with a clock frequency of 3.7~GHz, running Ubuntu~18.04.2 was used. This is the same hardware and software that was used in Reference~\cite{eligant-gg} where detector efficiencies were well reproduced up to an energy of 1.4~MeV. Due to the small amount of material present in the experimental area, see Figure~\ref{fig:setup}, we simplified the setup by only considering the active detector volume in the previously described geometry. The simulations were performed in steps of 5~keV, between 250~keV and 15~MeV, with 100~000 $\gamma$-rays emitted in the detector direction for each energy for a total of $295.1\cdot 10^{6}$ events and, approximately, seven days of computer time. The energy of the detected $\gamma$ rays were randomly shifted in energy based on a Gaussian distribution with the measured energy resolution for each event. The energy resolution was obtained from the experimental data using the same transitions as for the energy resolution and interpolated as
\begin{equation}
 \frac{\sigma_{E_{\gamma}}}{E_{\gamma}}  = \sqrt{\frac{1}{E_{\gamma}N_{\mathrm{phe}}}(1+\epsilon_{\mathrm{PMT}}) +\sigma_{\mathrm{noise}}^{2}},\label{eq:resolution}
\end{equation} 
based on the discussion in Reference~\cite{eres}, with fitted parameters $N_{\mathrm{phe}}=9.0$~keV$^{-1}$, $\epsilon_{\mathrm{PMT}}=0.20$, and $\sigma_{\mathrm{noise}} = 0.75$\%. These parameters can be roughly interpreted as the number of photo-electrons produced in the \ac{PMT} per keV of deposited energy ($N_{\mathrm{phe}}$), the variance of the \ac{PMT} ($\epsilon_{\mathrm{PMT}}$), and a generic  noise term ($\sigma_{\mathrm{noise}}$) originating from electronics and similar. Note that these values are just fitted to interpolate the energy resolution of the detector, and should not be considered reliable measurements of the physical quantities. They are, however, similar in magnitude to what we expect them to be in a dedicated evaluation. Finally, the spectrum of detected $\gamma$-rays were normalized to one for each emitted energy, to maintain the number of events in the spectrum after unfolding. 

\begin{figure}[ht]
 \centering
 \includegraphics[width=0.32\columnwidth]{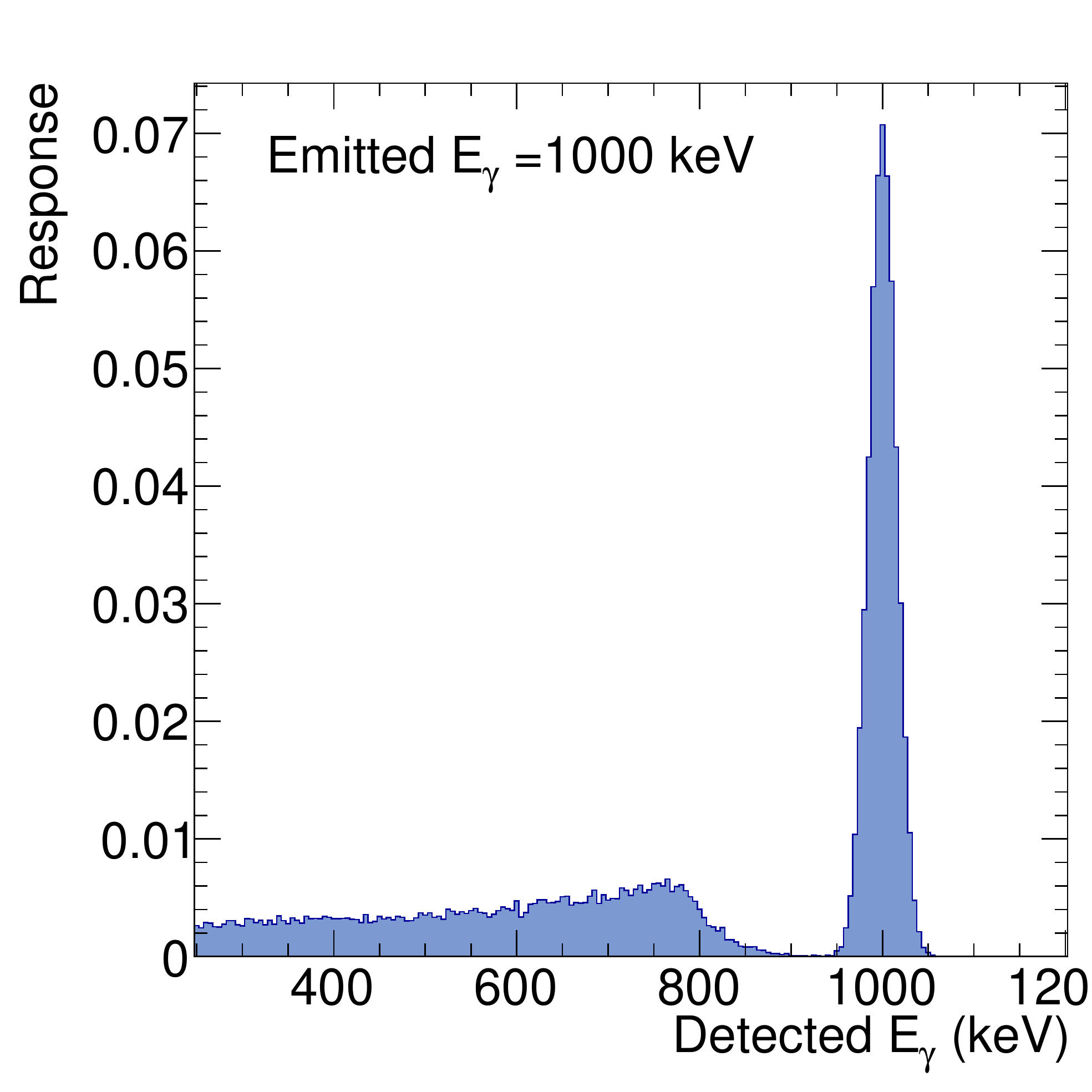}
 \includegraphics[width=0.32\columnwidth]{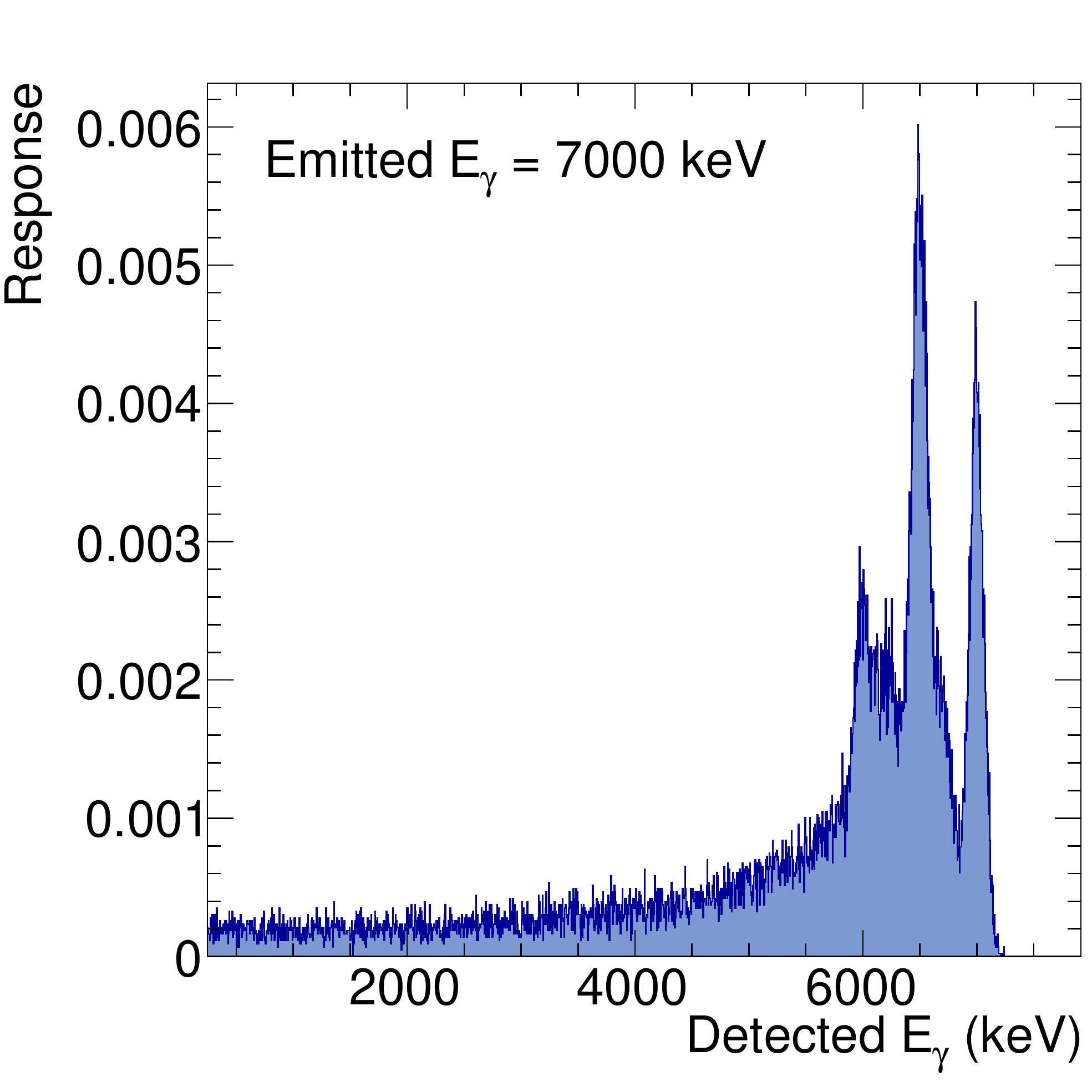}
 \includegraphics[width=0.32\columnwidth]{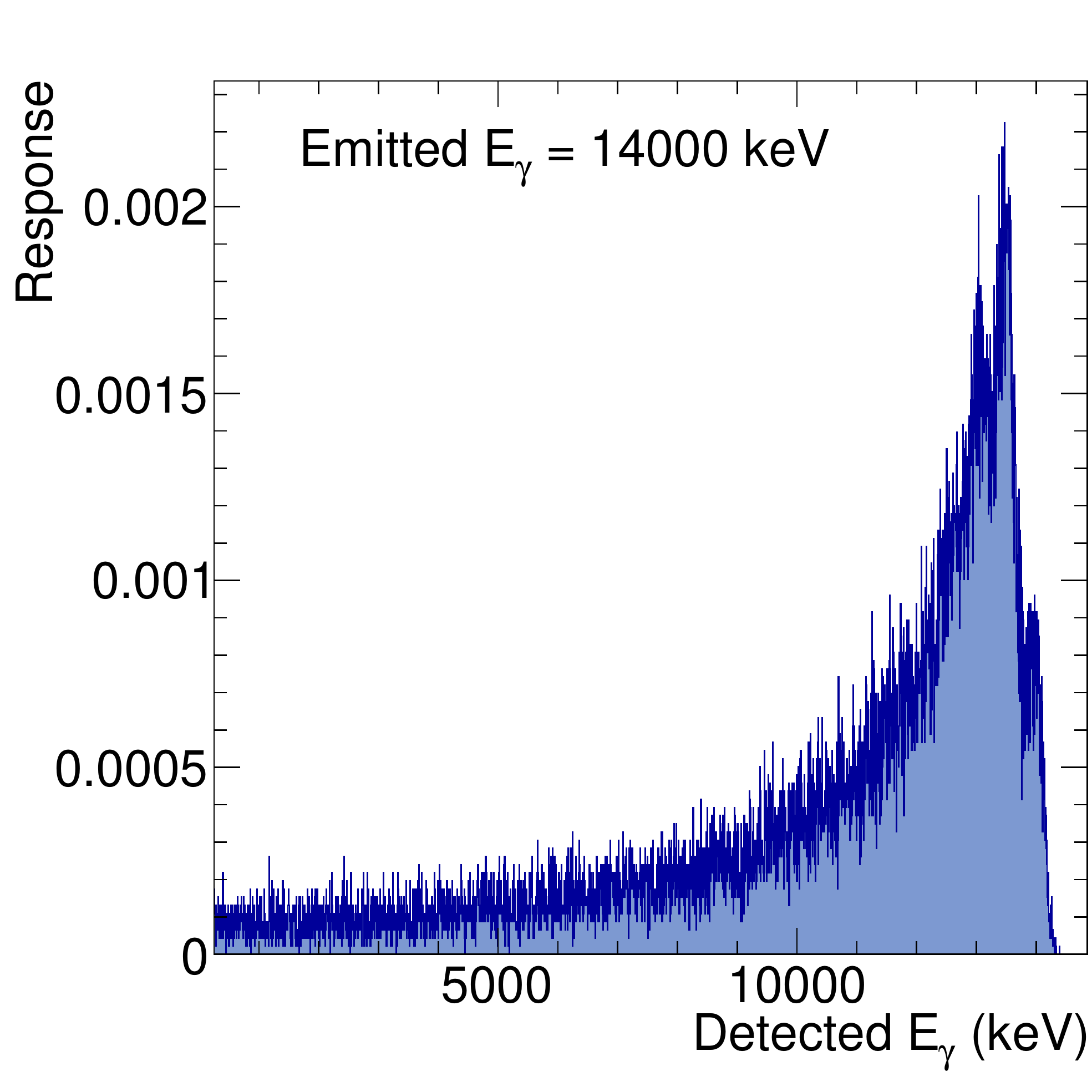}
 \caption{ Projection of the response matrix for emitted $\gamma$-ray energies of 1000~keV (left), 7000~keV (middle), and 14000~keV (right). \label{fig:response}}
\end{figure}

The projections of the response matrix at three selected energies are shown in Figure~\ref{fig:response}. As expected the full-energy peak is dominating the response matrix at low energies. At a few MeV $\gamma$-ray energy a significant part of the energy deposition is starting to go into the single escape peak and the Compton edge, while remaining distinguishable features. Above a $\gamma$-ray energy of approximately 10~MeV the energy deposition in the full-energy peak is strongly reduced and the measured spectrum is dominated by Compton scattering  with a minor fine structure coming from the escape peaks. Besides these features the projection of the spectra are rather smooth and featureless. This is due to the simplified geometry used in this work, motivated by the small amount of material that could induce secondary scattering in this setup. Another motivation for this simplification was that the low-energy thresholds in the data acquisition was at 200~keV, and the lower edge of the histograms used for the analysis at 250~keV. These are the energies where structures  originating from, for example, backscattering typically appear. Thus, this simplification is expected to only have minor influence at the low-energy edge of the acquired and subsequently unfolded histograms.

\section{The matrix inversion method}

\begin{figure}[ht]
 \centering
 \includegraphics[width=0.49\columnwidth]{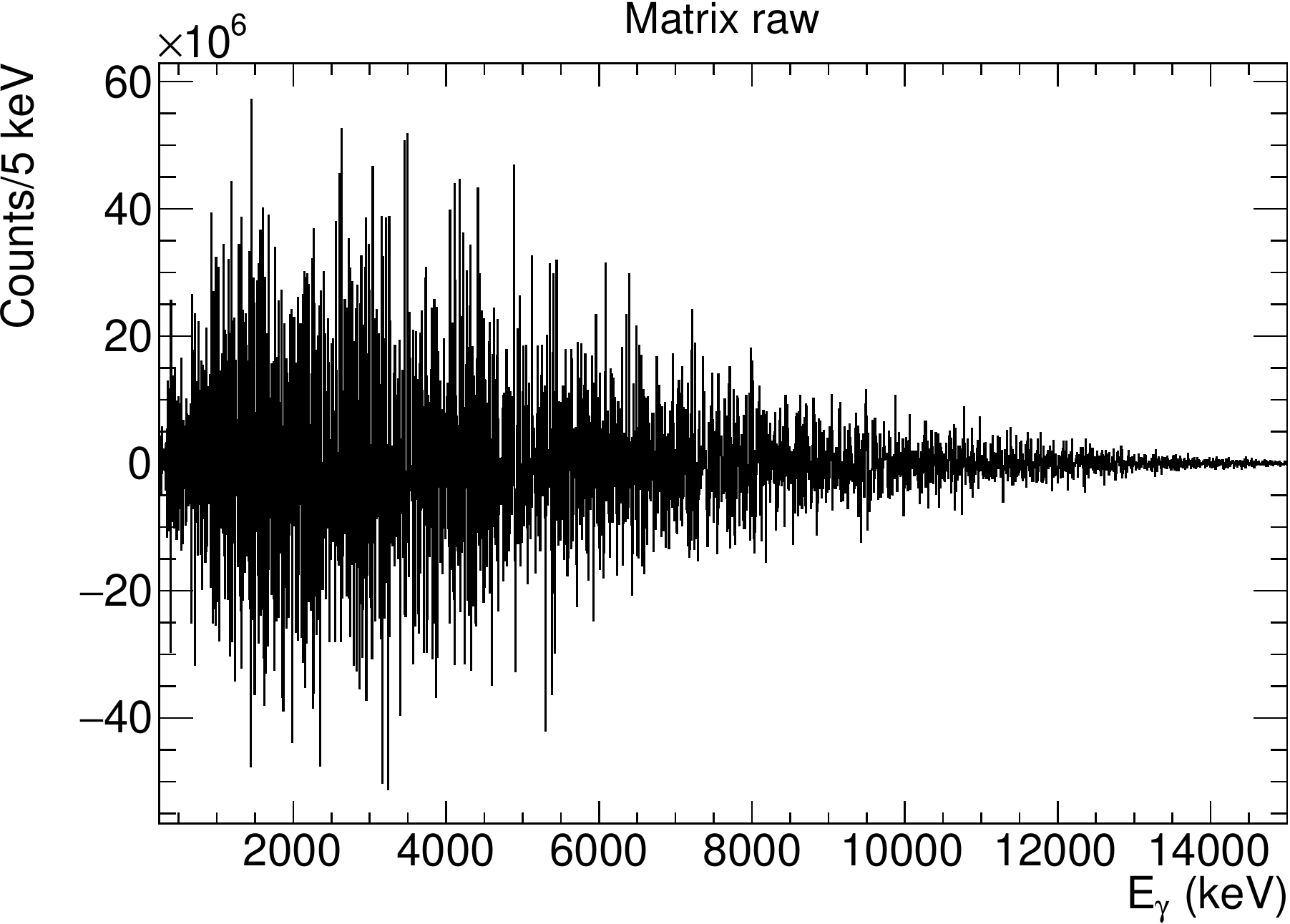}\\
 \includegraphics[width=0.49\columnwidth]{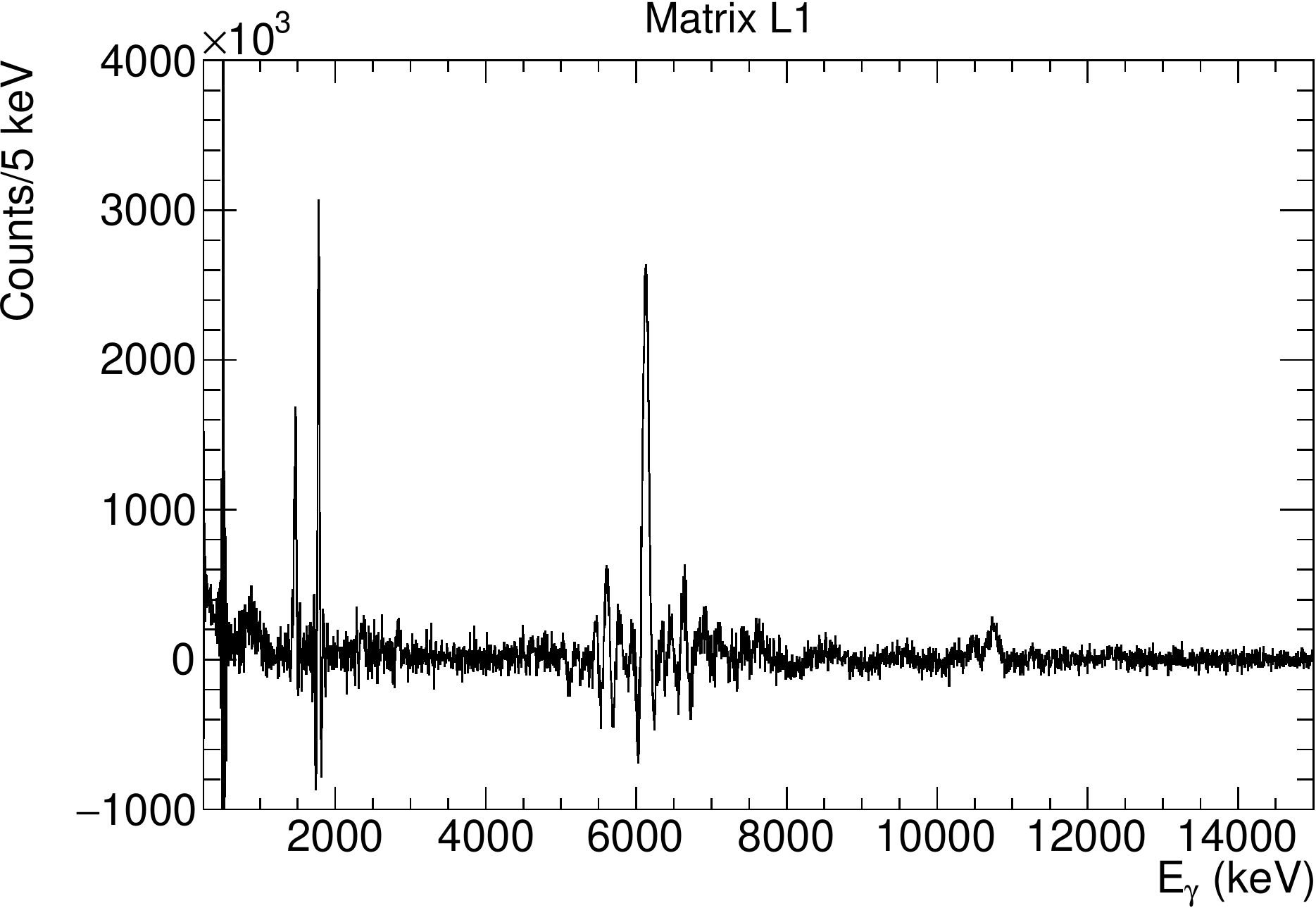}
 \includegraphics[width=0.49\columnwidth]{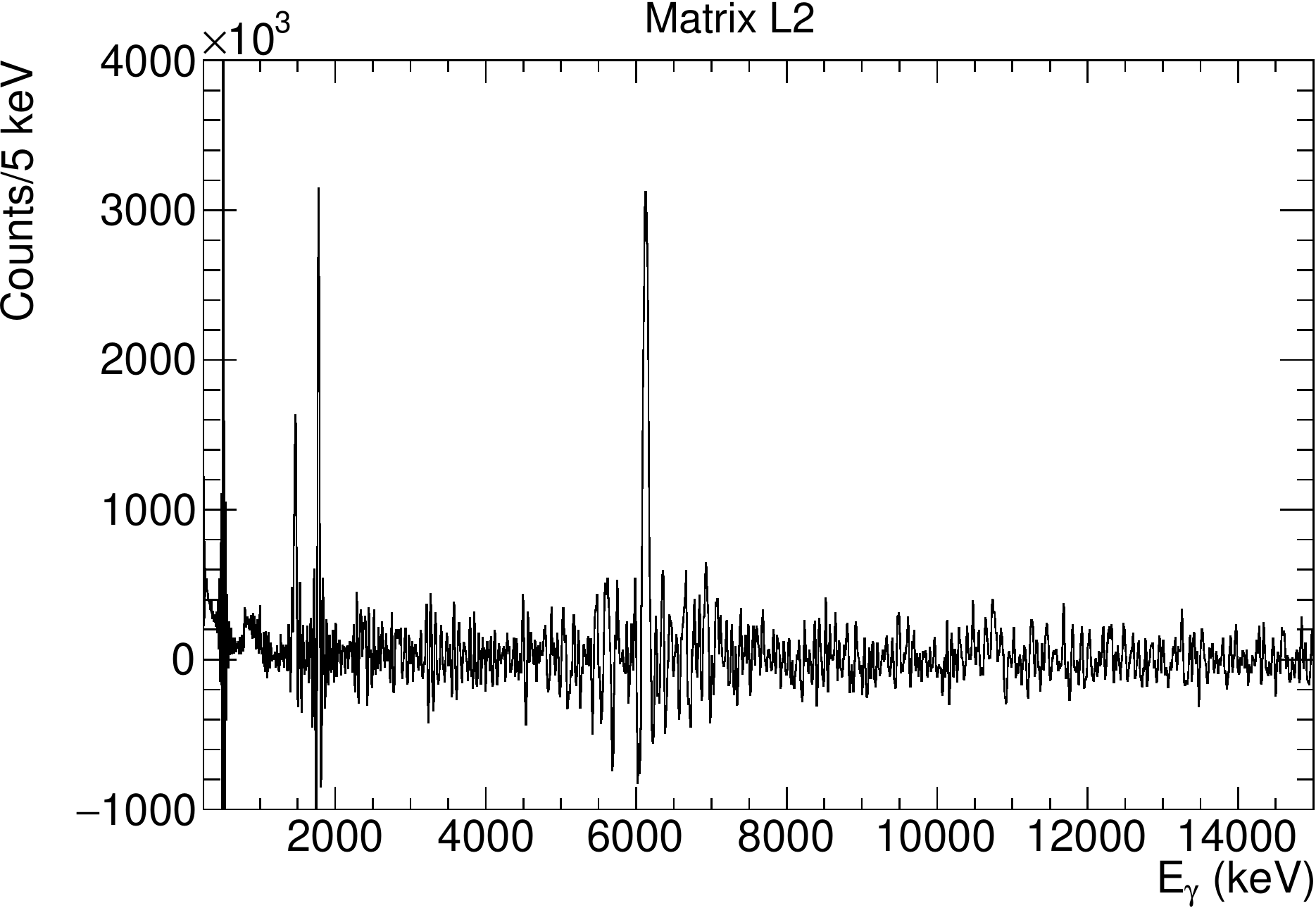}
 \caption{(Top) Unfolded energy spectrum using the matrix inversion method without regularization. (Bottom left) Same as the top spectrum but with Thikhonov-Phillips regularization of strength $\tau=10^{-4}$. (Bottom right) Same as the top spectrum but with second derivative regularization of strength $\tau=10^{-4}$  \label{fig:matrix_noreg}}
\end{figure}

The most straightforward way to obtain the true spectrum, $\vec{x}$, given a measured spectrum, $\vec{y}$, and a response matrix $\mathbf{A}$ is by the matrix inversion method \cite{matrix_ref_1,matrix_ref_2},
\begin{equation}
 \vec{x} = \mathbf{A}^{-1}\vec{y},
\end{equation} 
where we define the relation between covariance matrices $\mathbf{V}_{x,y}$  as
\begin{equation}
 \mathbf{V}_{x} = \mathbf{A}^{-1}\mathbf{V}_{y}\left(\mathbf{A}^{-1}\right)^{\mathrm{T}}.
\end{equation} 
In the following discussion, $\mathbf{V}_{y}$ is assumed to be the identity matrix. While this method is, in principle, perfect, the mathematical nature of inverse matrices will introduce large alternating positive and negative values in $\mathbf{A^{-1}}$ due to very strong negative correlations in $\mathbf{V}_{x}$. Thus, while the unfolded spectrum, shown in the top section of \tfig\ref{fig:matrix_noreg}, from a statistical point of view is completely correct it is not very useful for evaluation of  the experimental data. While not a topic within the scope of this report, note that these large negative and positive values, although unphysical, contain important correlations in the covariance matrix and need to be kept for any propagation of uncertainties to be correct, for the methods evaluated here.

One common solution to this is to make the solution less perfect by introducing a regularization parameter, $\tau$, that smooths out $\mathbf{A^{-1}}$ and reduces the correlations of $\mathbf{V}_{x}$ \cite{blobel}. There are several methods to do this, but in this work we will look at two simple cases. The first is the Thikhonov-Phillips regularization \cite{Tikhonov,Phillips} where $\tau$ is introduced as a constant to the identity matrix, $\mathbf{1}$, as
\begin{equation}
 \mathbf{A}^{\mathrm{T}}\mathbf{V}_{y}^{-1}\vec{y} = \left(\mathbf{A}^{\mathrm{T}}\mathbf{V}_{y}^{-1}\mathbf{A}+\tau\mathbf{1}\right)\vec{x}.
\end{equation} 
This will smooth the solution with respect to $\vec{x}$, as shown in the bottom left of \tfig\ref{fig:matrix_noreg}. We will refer to this method as Matrix~L1 hereafter. The other approach is to smooth the solution with respect to the second derivative of $\vec{x}$ \cite{blobel} using the matrix $\mathbf{L}$ of size $n \times m+2$ as,
\begin{equation}
 \mathbf{A}^{\mathrm{T}}\mathbf{V}_{y}^{-1}\vec{y} = \left(\mathbf{A}^{\mathrm{T}}\mathbf{V}_{y}^{-1}\mathbf{A}+\tau\mathbf{L}^{\mathrm{T}}\mathbf{L}\right)\vec{x}.\label{eq:snd_der}
\end{equation} 
In this case $\mathbf{L}$ is constructed so that for each row $\vec{L}_{n}=(0,\ldots,L_{n,m}=1,L_{n,m+1}=-2,L_{n,m+2}=1,0,\ldots)$. Applying equation~(\ref{eq:snd_der}) to $\mathbf{A}$ and $\vec{y}$ from \tfig\ref{fig:setup}, with $\tau=10^{-4}$, produces the  emission spectrum, $\vec{x}$, shown in the bottom right of \tfig\ref{fig:matrix_noreg}. We will refer to this method as Matrix~L2 hereafter.

\section{The iterative unfolding procedure}

Two major drawbacks of the matrix inversion procedure are the need for regularization, and the very computationally intense processes of large matrix inversion. To avoid these issues the iterative unfolding procedure \cite{iterative_1,iterative_2} has been implemented within the framework of the Oslo method for experiments on level densities and $\gamma$ strength functions
\cite{Guttormsen1987,Guttormsen1996,Larsen2011}. A detailed discussion about the possible systematic errors in the Oslo method can be found in \tref\cite{Larsen2011}. This method has typically been used successfully for unfolding of high $\gamma$-ray density spectra with $\gamma$-ray energies up to 7~MeV \cite{PhysRevC.73.064301,PhysRevC.76.044303,PhysRevC.79.024316} with some examples up to 10~MeV \cite{PhysRevC.83.014312}. In \tref\cite{Guttormsen1996} the method is evaluated for discrete $\gamma$-ray spectra following the decay of $^{152}$Eu up to $\gamma$-ray energies of 1.4~MeV.

\begin{figure}[ht]
 \centering
 \includegraphics[width=0.49\columnwidth]{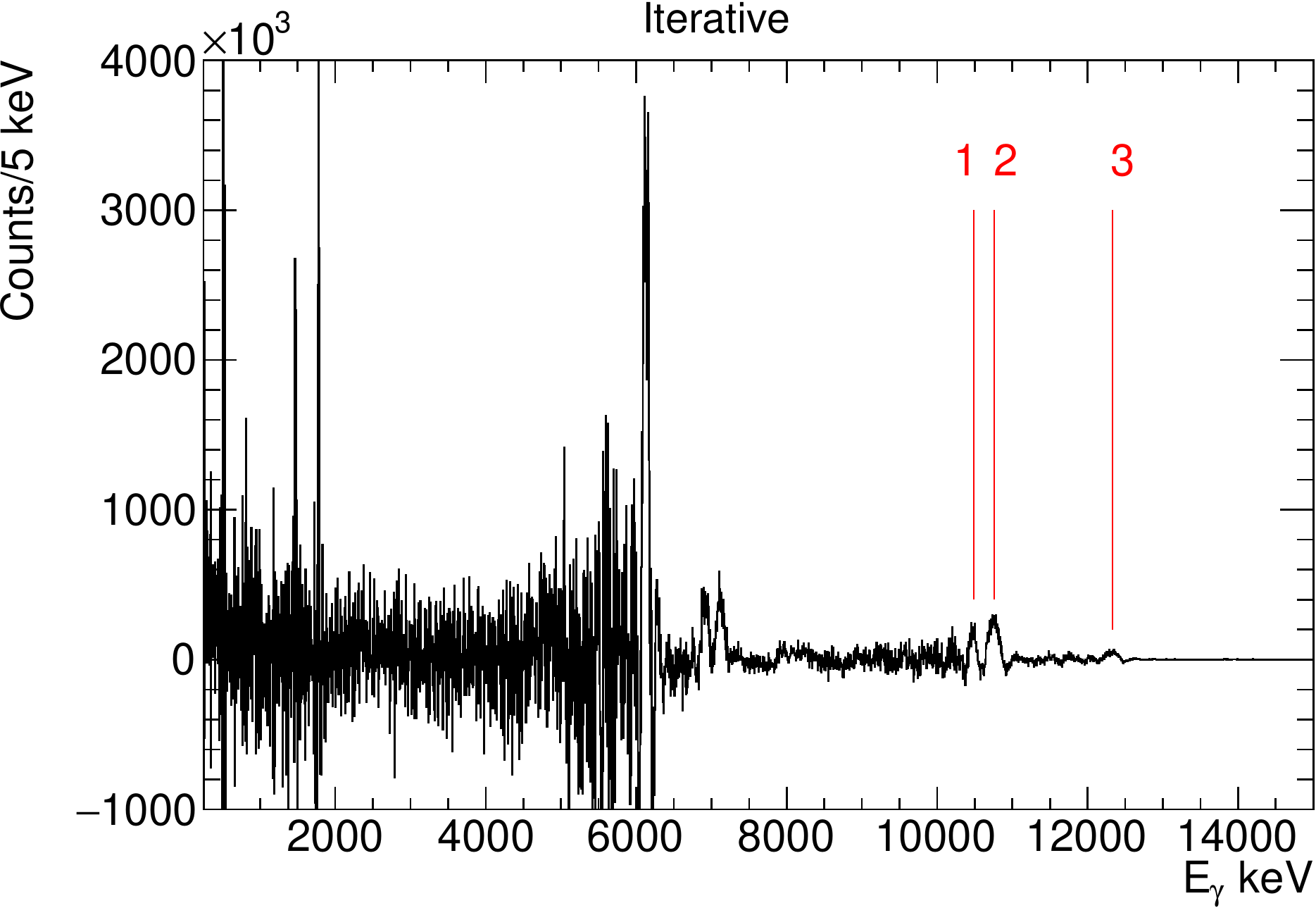}
 \includegraphics[width=0.49\columnwidth]{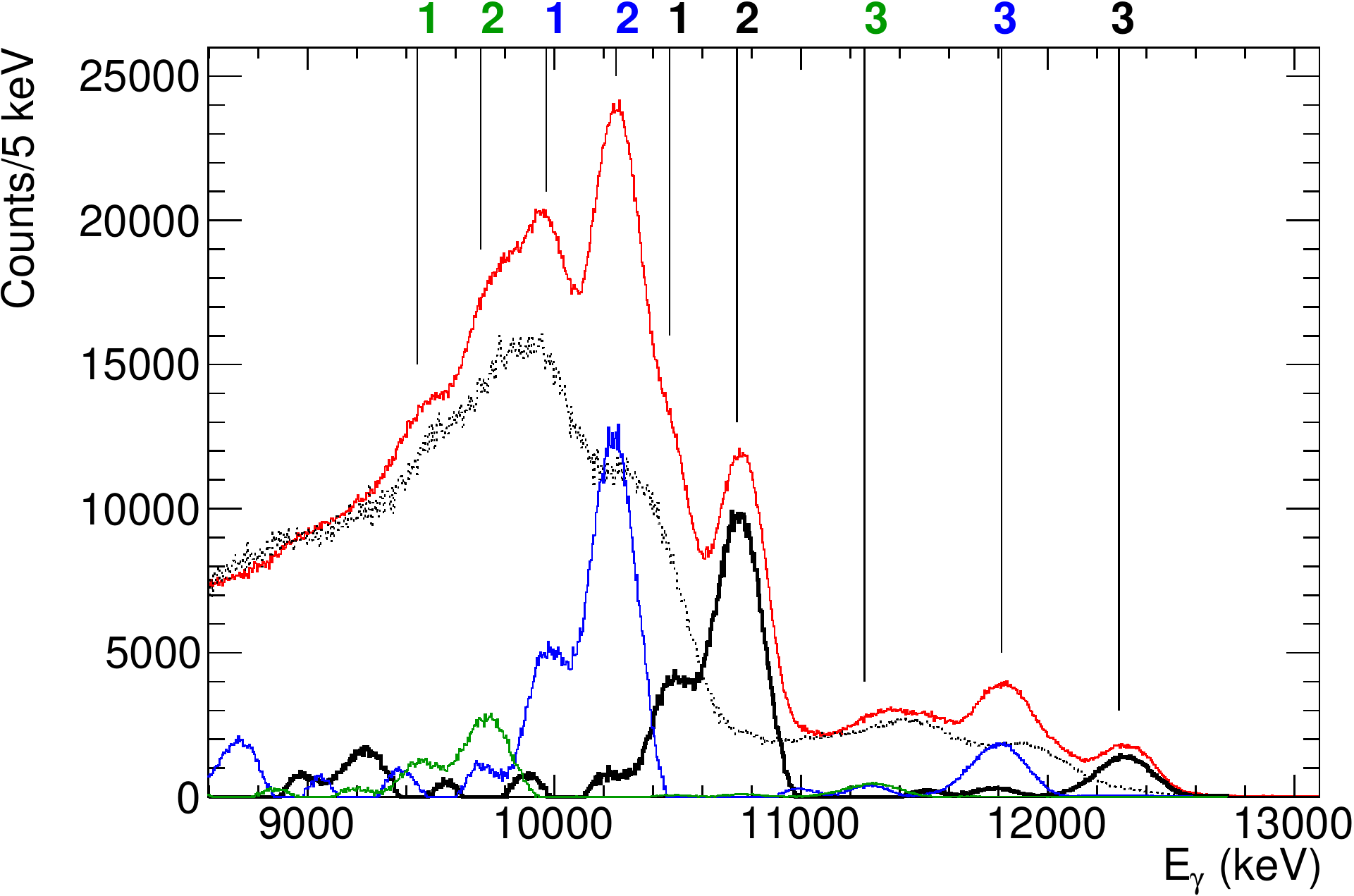}
 \caption{(Left) Unfolded energy spectrum using the iterative approach. (Right) Measured raw spectrum in the high-energy regime (red) and decomposed into the full-energy component (solid black), Compton component (dotted black), single-escape events (blue) and double-escape events (green). Three peaks have been identified in this spectrum marked with 1, 2, and 3, respectively. \label{fig:oslo_unfolded}}
\end{figure}

The iterative unfolding procedure is based on successive refolding of a starting guess spectrum. We call the unfolded spectrum in an iteration $i$, $\vec{x}{'}_{i}$ and a refolded spectrum $\vec{y}{'}_{i}$. For a starting guess, usually $\vec{x}{'}_{0}=\vec{y}$ can be used. This gives the first step of the procedure as
\begin{equation}
 \vec{y}{'}_{0} = \mathbf{A}\vec{x}{'}_{0} = \mathbf{A}\vec{y}.
\end{equation} 
In the next iteration the starting guess is modified as
\begin{equation}
 \vec{x}{'}_{1} = \vec{x}{'}_{0} + \left(\vec{y}-\vec{y}{'}_{0}\right),
\end{equation} 
and the procedure is repeated for each iteration $i$
\begin{equation}
\begin{split}
 \vec{x}{'}_{i} &= \vec{x}{'}_{i-1} + \left(\vec{y}-\vec{y}{'}_{i-1}\right),\\
 \vec{y}{'}_{i} &= \mathbf{A}\vec{x}{'}_{i},
\end{split} 
\end{equation} 
until convergence. Typically the iterative process is repeated tens of times before it converges. In this work, however, since we are mainly interested in the high-energy response, a larger number of iterations was necessary before convergence of the high-energy part. Thus, a Kolmogorov-Smirnov test \cite{kolmogorov,smirnov} with a cut-off at a Kolmogorov similarity of $\alpha=99.73$\%, corresponding to 3$\sigma$, between the original and unfolded-refolded spectra was employed to define convergence. Using $\mathbf{A}$ and $\vec{y}$ from \tfig\ref{fig:setup}, the result after 197 iterations is shown in \tfig\ref{fig:oslo_unfolded}.

The existence of two peak-like structures around 10~MeV can clearly be seen in the unfolded spectrum. In \tfig\ref{fig:oslo_unfolded} we show the spectrum in this region with a decomposition into the full-energy peak, the single- and double-escape peaks, and the Compton continuum. We see that the second peak is hidden under the distribution from the larger peak at slightly higher energy in the raw spectrum, clearly identified in the decomposed distributions.

\section{Discussion}

\begin{figure*}[ht]
 \centering
 \includegraphics[width=0.32\textwidth]{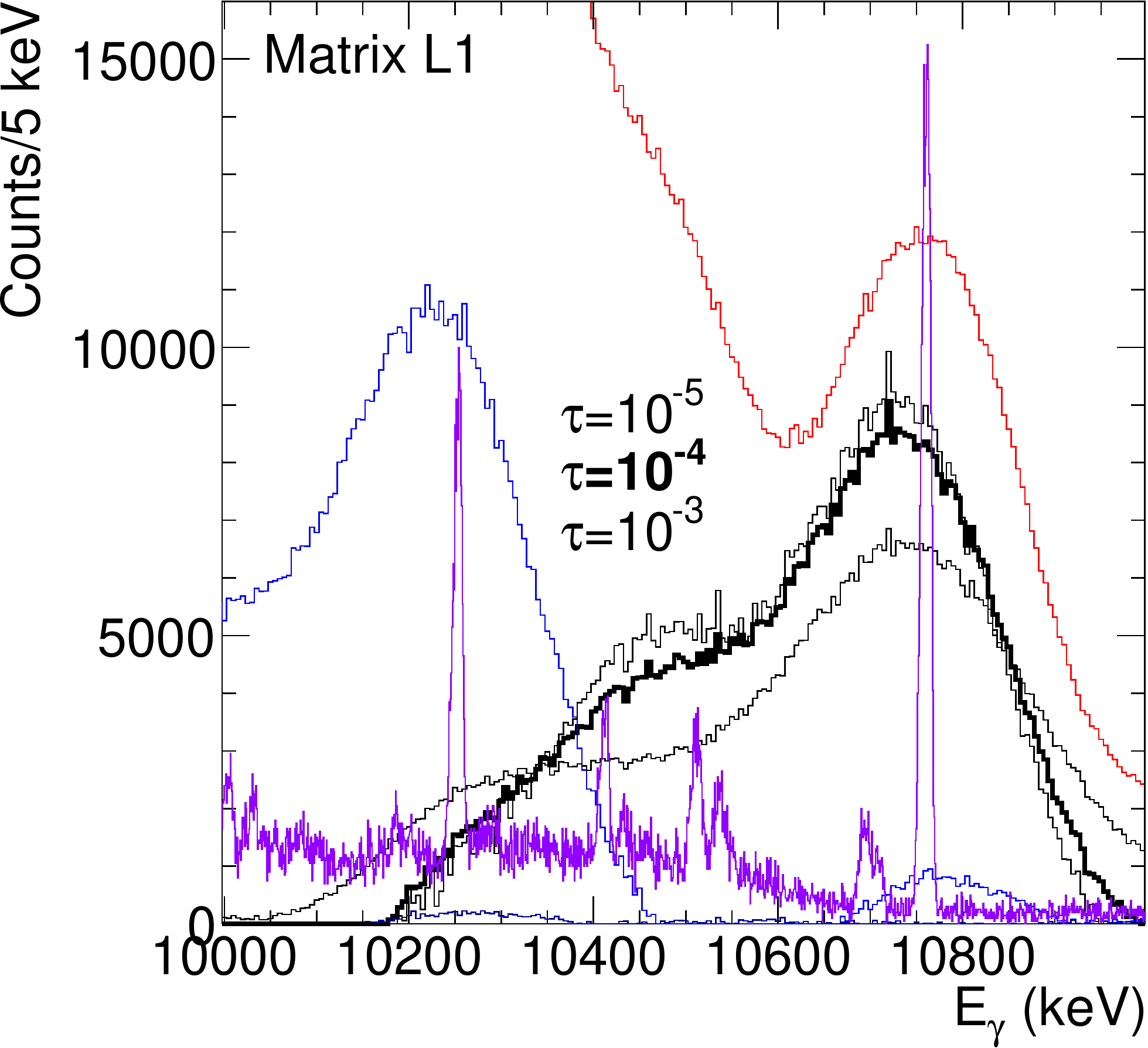}
 \includegraphics[width=0.32\textwidth]{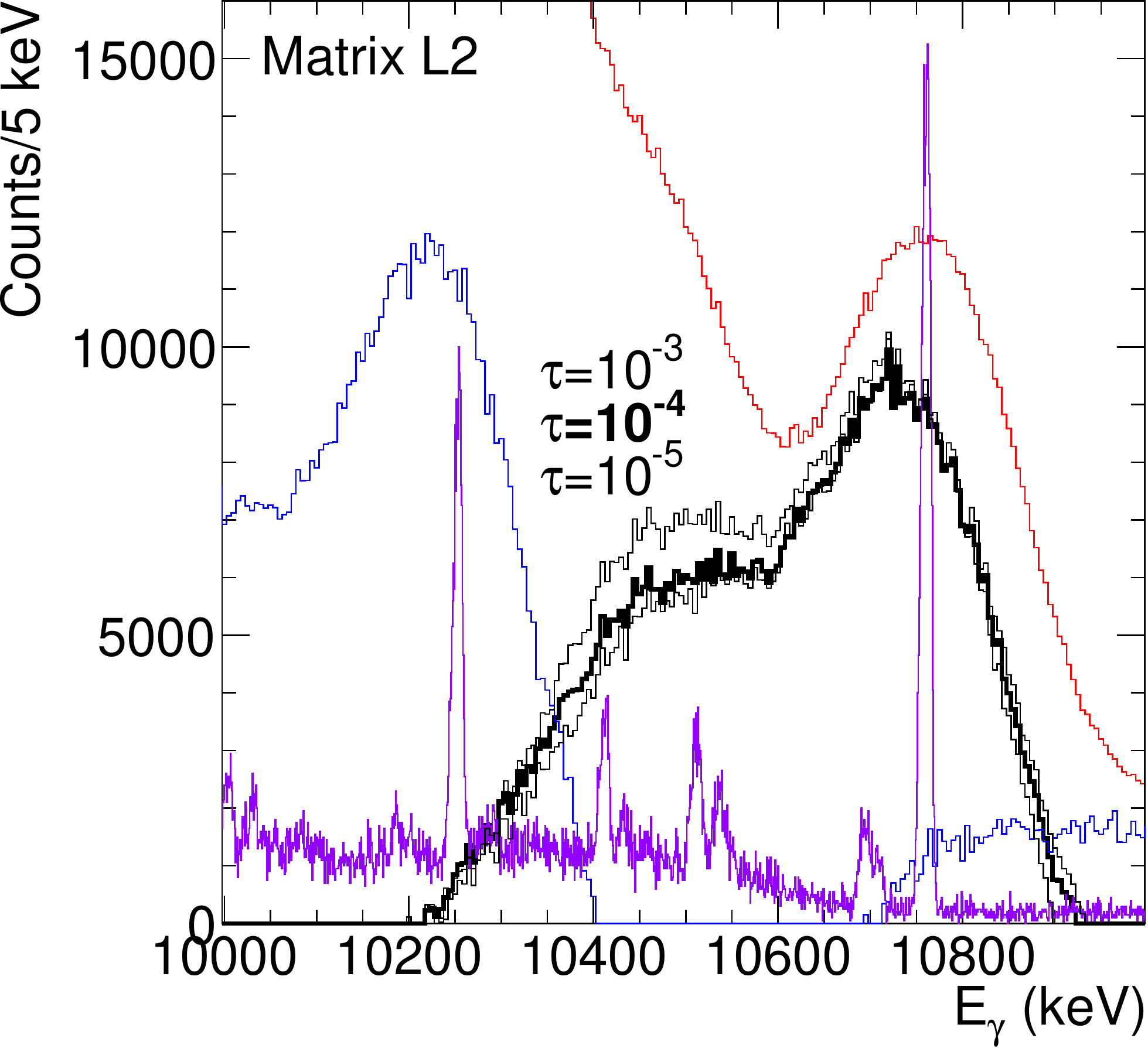}
 \includegraphics[width=0.32\textwidth]{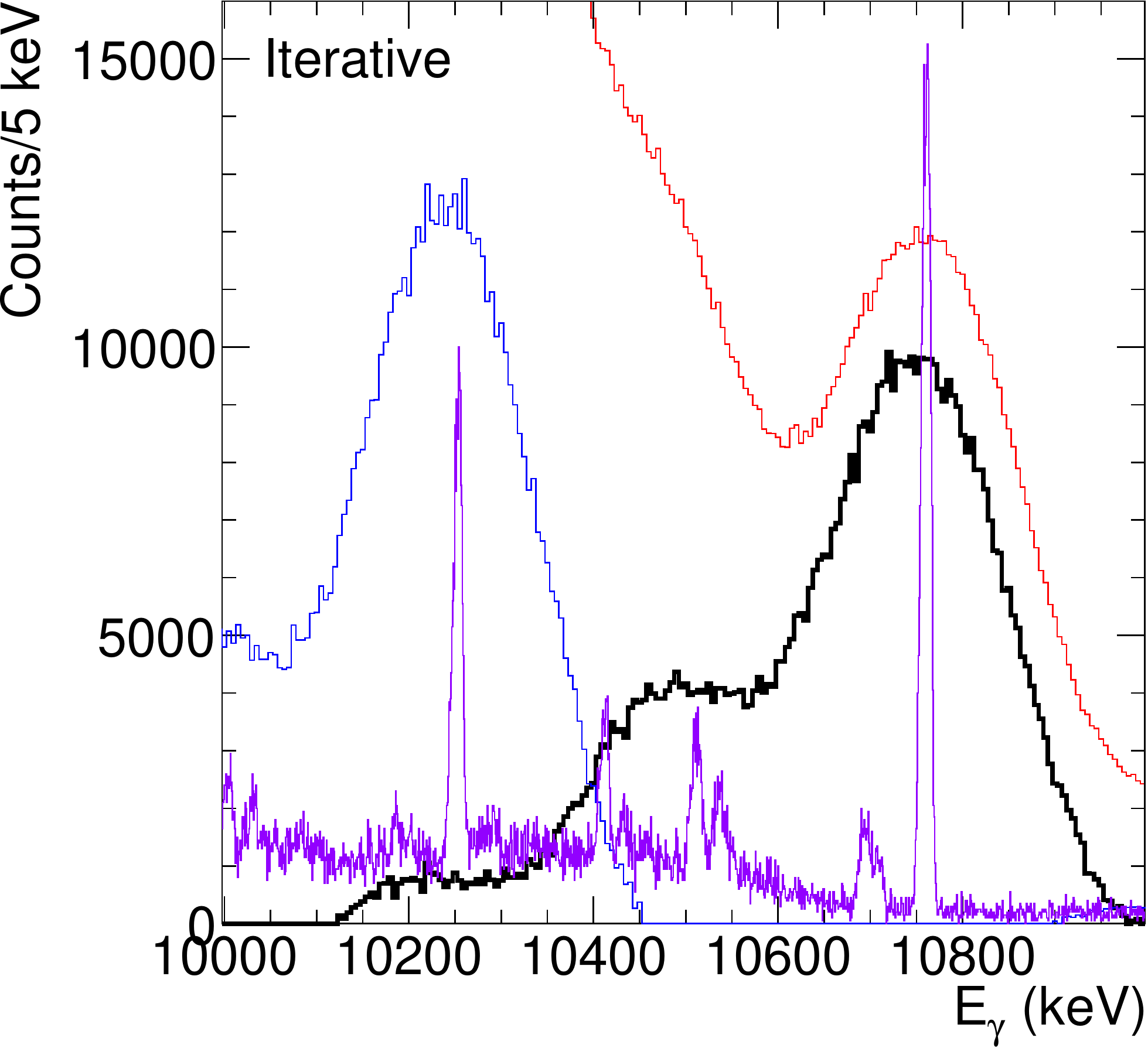}\\
 \includegraphics[width=0.32\textwidth]{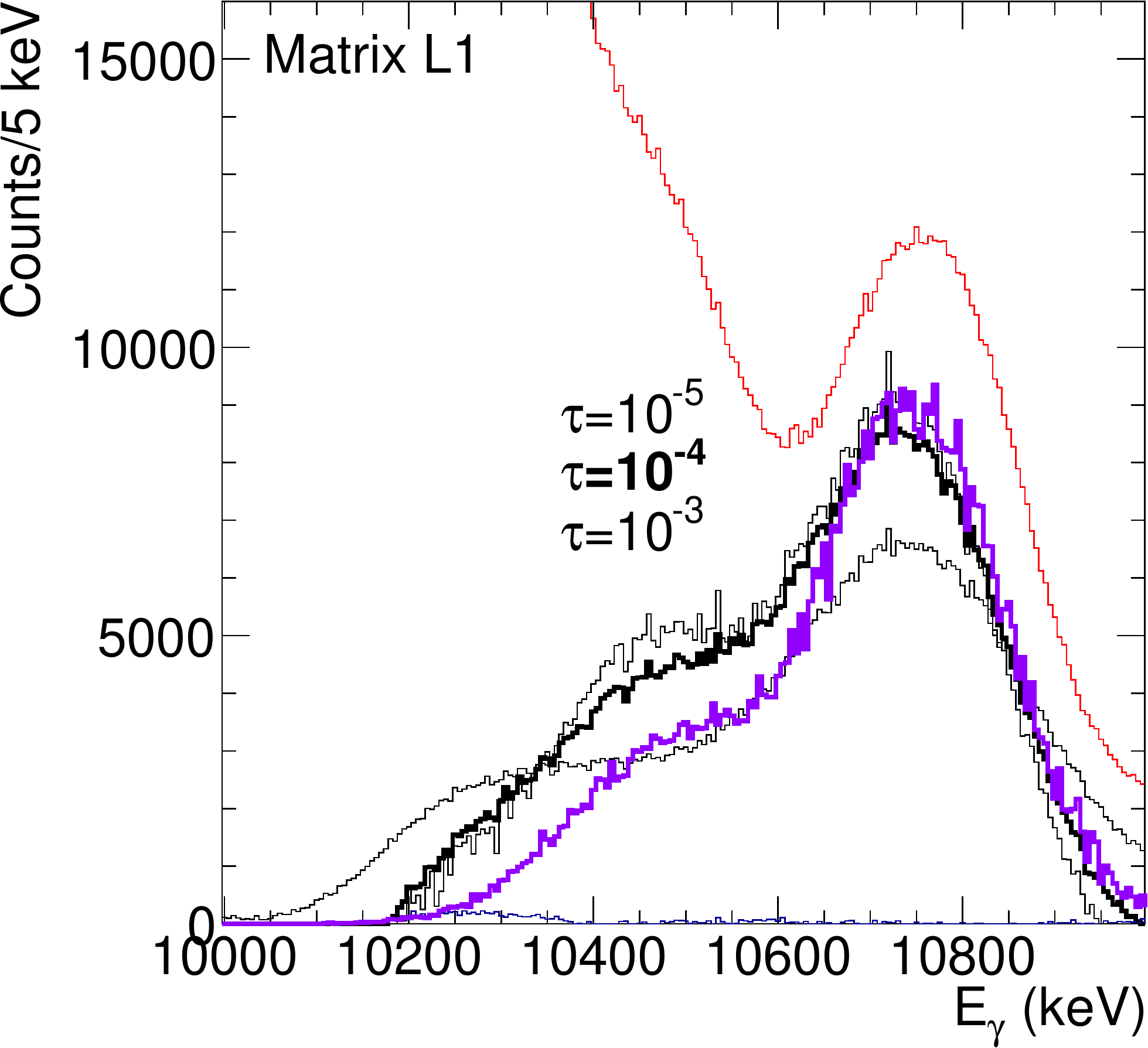}
 \includegraphics[width=0.32\textwidth]{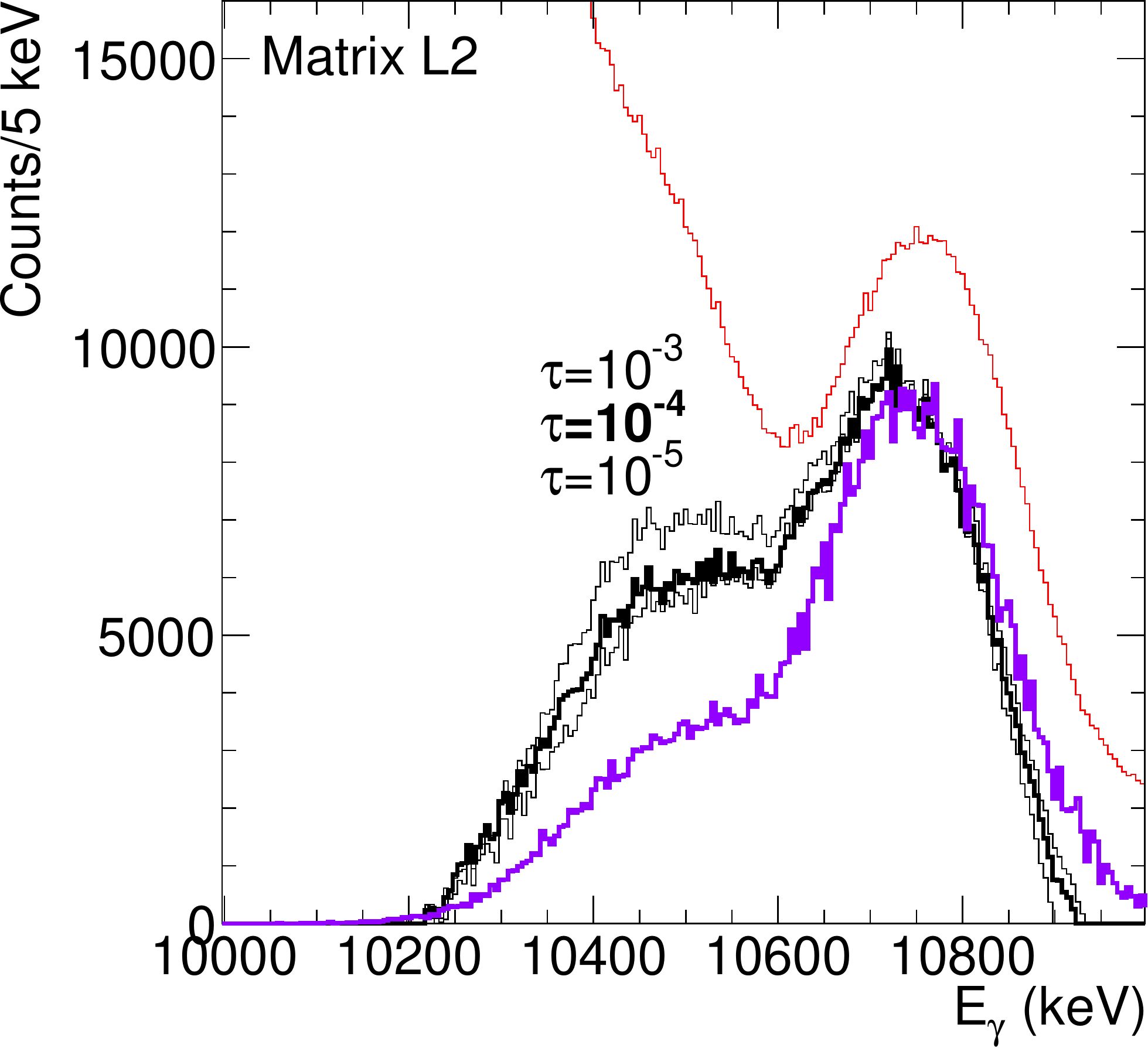}
 \includegraphics[width=0.32\textwidth]{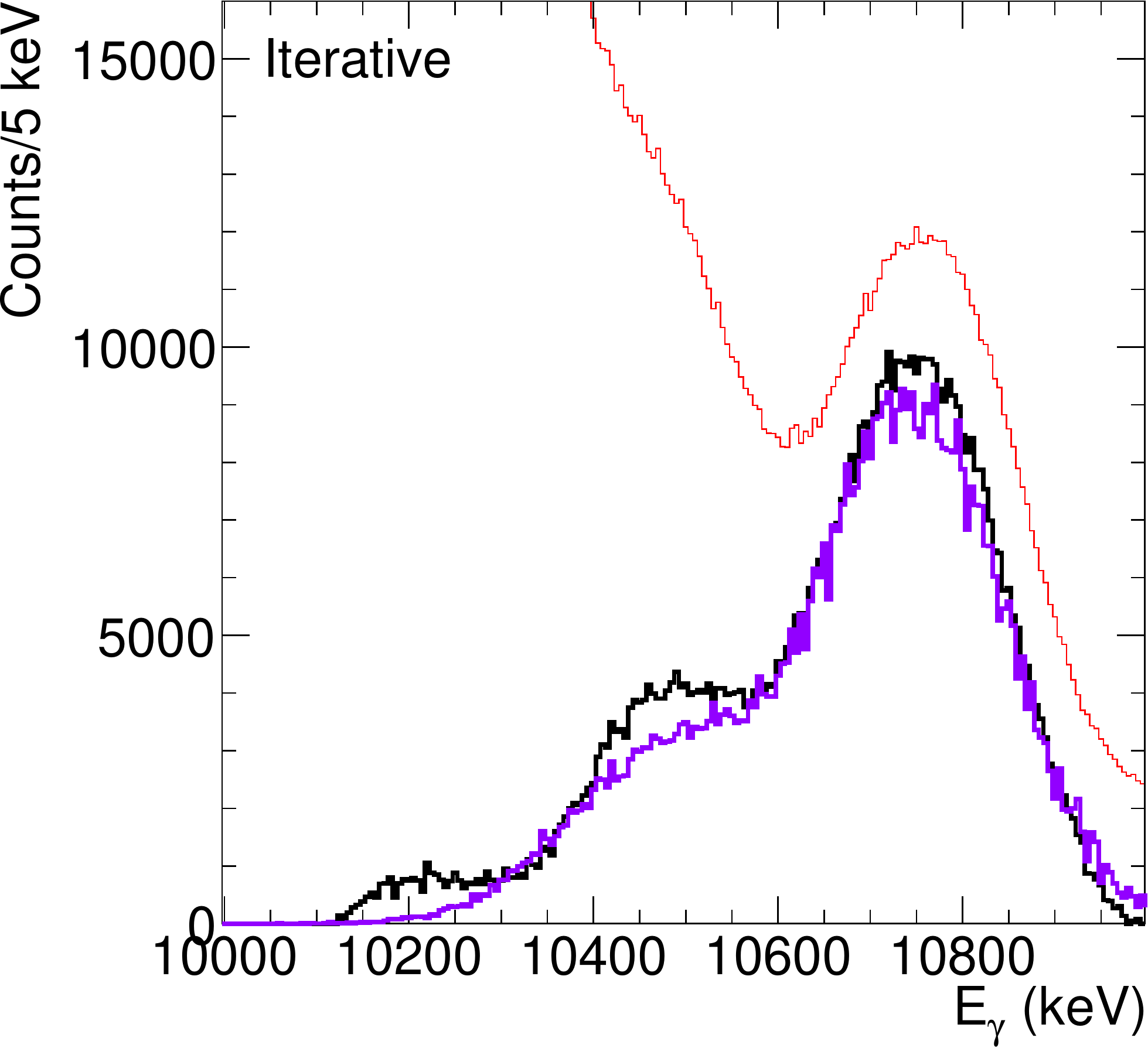}\\
 \caption{Raw energy (red), full-energy deposition (black), and single-escape (blue) spectra in the high energy regime for the Thikhonov-Phillips matrix regularization (Matrix L1, left), the second derivative regularization (Matrix L2, middle) and the iterative unfolding procedure (right). The purple spectrum shows HPGe data obtained in similar experimental conditions. The bottom row is the same as the top row, but here the full-energy peaks in the HPGe spectrum have been refolded with the \labr\ full-energy peak response and renormalized to the same intensity.\label{fig:zoom}}
\end{figure*}

To qualitatively estimate the performance of the different unfolding algorithms, the same decomposition as in \tfig\ref{fig:oslo_unfolded} into different components was performed for all three unfolding methods. To verify that the observed hidden structure is not only an artifact from the unfolding algorithms, or imperfections in the energy calibration of the experimental data, we compare the results to data obtained by a \ac{HPGe} detector under the same experimental conditions. The results from these decompositions are shown in \tfig\ref{fig:zoom}. It is clear from the \ac{HPGe} spectrum that there are, indeed, two minor structures from other silicon resonances, close in energy with a similar intensity as obtained by the \labr\ spectrum at this energy.

Comparing the three different methods qualitatively we note from \tfig\ref{fig:zoom} that all of them identify the small structure at lower energies. However, we also note that the Thikhonov-Phillips regularization scheme underestimates the height of the peaks while giving a broader distribution. This is not unexpected as the regularization is performed by smoothing $\vec{x}$ directly. The second-derivative regularization more accurately reproduces the width and the height of the main peak. Both of these schemes, however, significantly overestimate the size of the hidden lower-energy structure compared to what was obtained from the \ac{HPGe} data, as well as dropping below zero at the tails of the distributions. Aside from not reproducing the expected slight shift of the peak towards lower energies, something that both matrix inversion methods do, the best qualitative agreement between the \ac{HPGe} and the \labr\ data is given by the iterative approach. Here the hidden features are well reproduced in magnitude, although giving a slightly too prominent peak, something that is also observed in the second-derivative regularization. 
Furthermore, there is a sensitivity in the results from the matrix algorithms with respect to the choice of $\tau$, in particular for the Thikhonov-Phillips regularization, that is not present in the parameter-free iterative approach.

\begin{table*}[ht]
	\caption{Kolmogorov similarity (KS) and root-mean-square (RMS) values in the energy ranges 3-5~MeV, 8-10~MeV, and 13-15~MeV for the three different unfolding methods evaluated in this work.}
	\label{tab:parameters}
	\centering
		\begin{tabular*}{\textwidth}{@{\extracolsep{\fill}}|ll|cccc|}
			\hline
          Method  & & KS & RMS 3-5 & RMS 8-10 & RMS 13-15 \\
			\hline
    Unregularized &  &  & 820000 & 240000 & 26000 \\
    Matrix L1 & $\tau=10^{-6}$       & 1.0000 & 21000 & 16000 & 7600 \\
    & $\tau=10^{-5}$       & 0.99998 & 10000 & 7800 & 5000 \\
    & $\tau=10^{-4}$       & 0.99975 & 3400 & 3000 & 1800 \\
    & $\tau=10^{-3}$       & 0.99429 & 1600 & 1100 & 500 \\
    Matrix L2& $\tau=10^{-6}$        & 1.00000 & 17000 & 13000 & 7100 \\
    & $\tau=10^{-5}$        & 0.99998 &  10000& 8900 & 5800 \\
    & $\tau=10^{-4}$        & 0.99995 & 7400 & 6500 & 5500 \\
    & $\tau=10^{-3}$        & 0.99964 & 5400 & 5700 & 5500 \\
    Iterative &          & 0.99731 & 14000 &  2500& 50 \\
			\hline
\end{tabular*}
\end{table*}

A more quantitative comparison of the quality between the different unfolding methods can be obtained by examining the size of the positive-negative fluctuations in the spectra originating from the negative correlations in the unfolding procedure. This comparison is done using the \ac{RMS} of the unfolded spectrum in the intermediate energy ranges where we do not expect any significant peaks. The \ac{RMS} values for all three methods are listed in \ttab\ref{tab:parameters}. For this particular type of spectrum, the fluctuations are more evenly distributed in the matrix unfolding approach, while in the iterative approach the fluctuations are pushed towards lower energies. The size of the fluctuations for the low-energy part of the spectra using the iterative approach are about four and two times the size of the fluctuations from the Thikhonov-Phillips matrix regularization and the second-derivative regularization, respectively. In the intermediate energy range all the three methods give fluctuations of similar magnitude. For the high-energy range the fluctuations are almost two orders of magnitude smaller with the iterative approach compared to the second derivative matrix regularization approach.
It is interesting to note that the Kolmogorov similarity, listed in \ttab\ref{tab:parameters}, is significantly lower for the spectrum unfolded with the iterative approach than the ones unfolded with the matrix approach, where, despite the rather large choice of $\tau$, both of the regularization schemes give very high Kolmogorov similarity. Thus, while the Kolmogorov similarity is a good indicator of the general quality of the result, it does not guarantee that details of the underlying spectra are well reproduced, as shown in Figure~\ref{fig:zoom}.

\begin{figure*}[ht]
 \centering
 \includegraphics[width=0.33\textwidth]{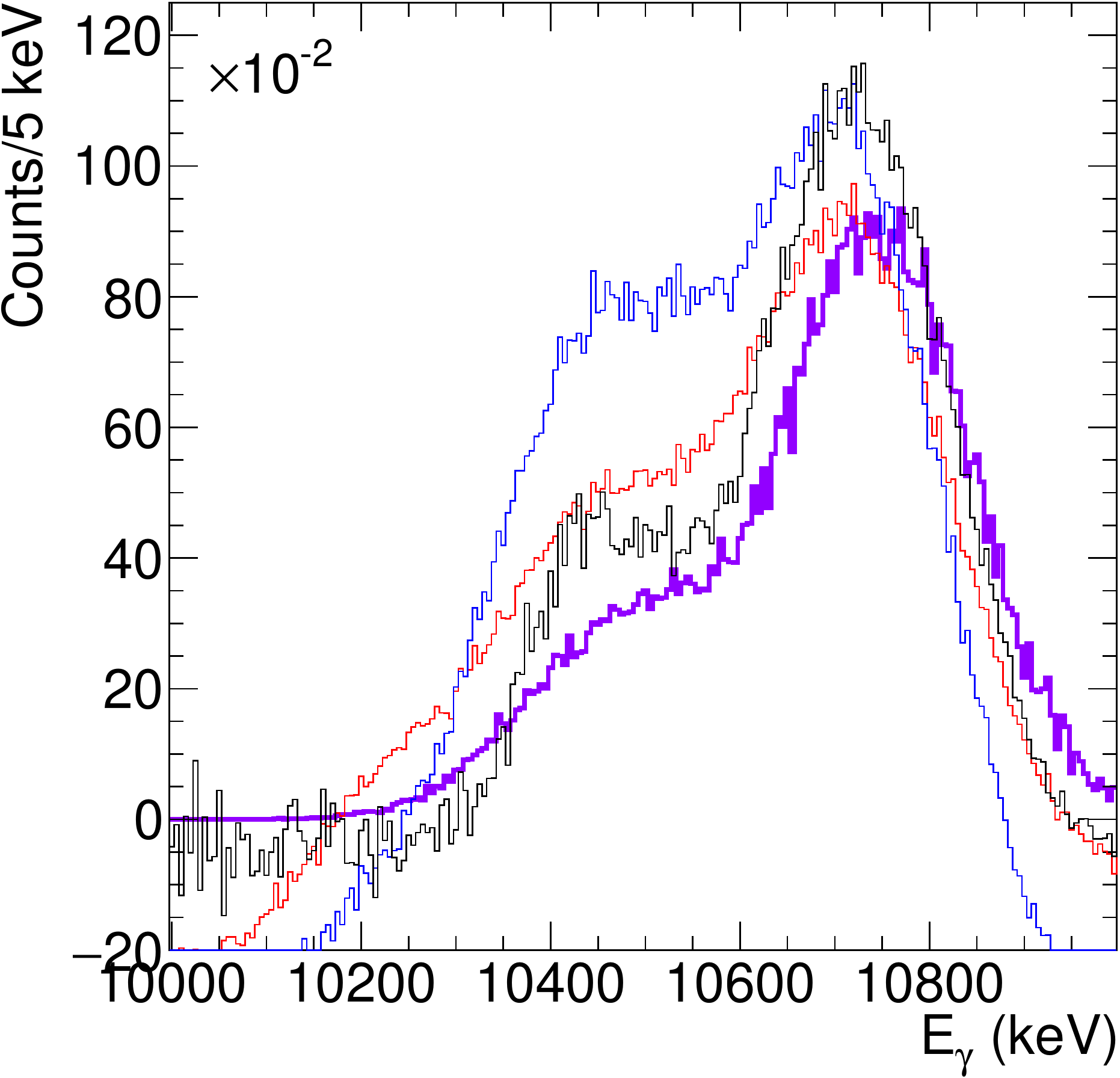}
 \includegraphics[width=0.32\textwidth]{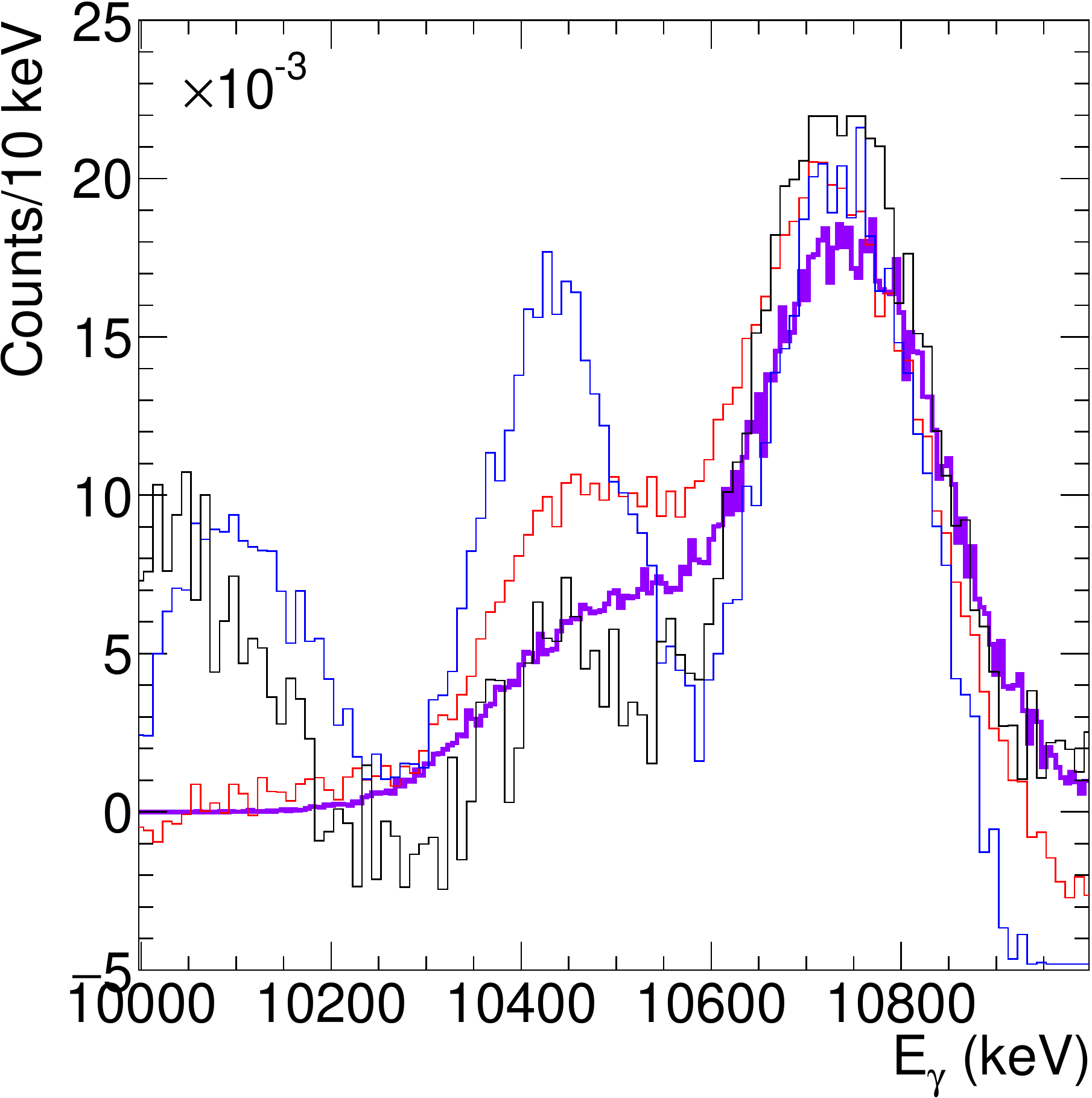}
 \includegraphics[width=0.31\textwidth]{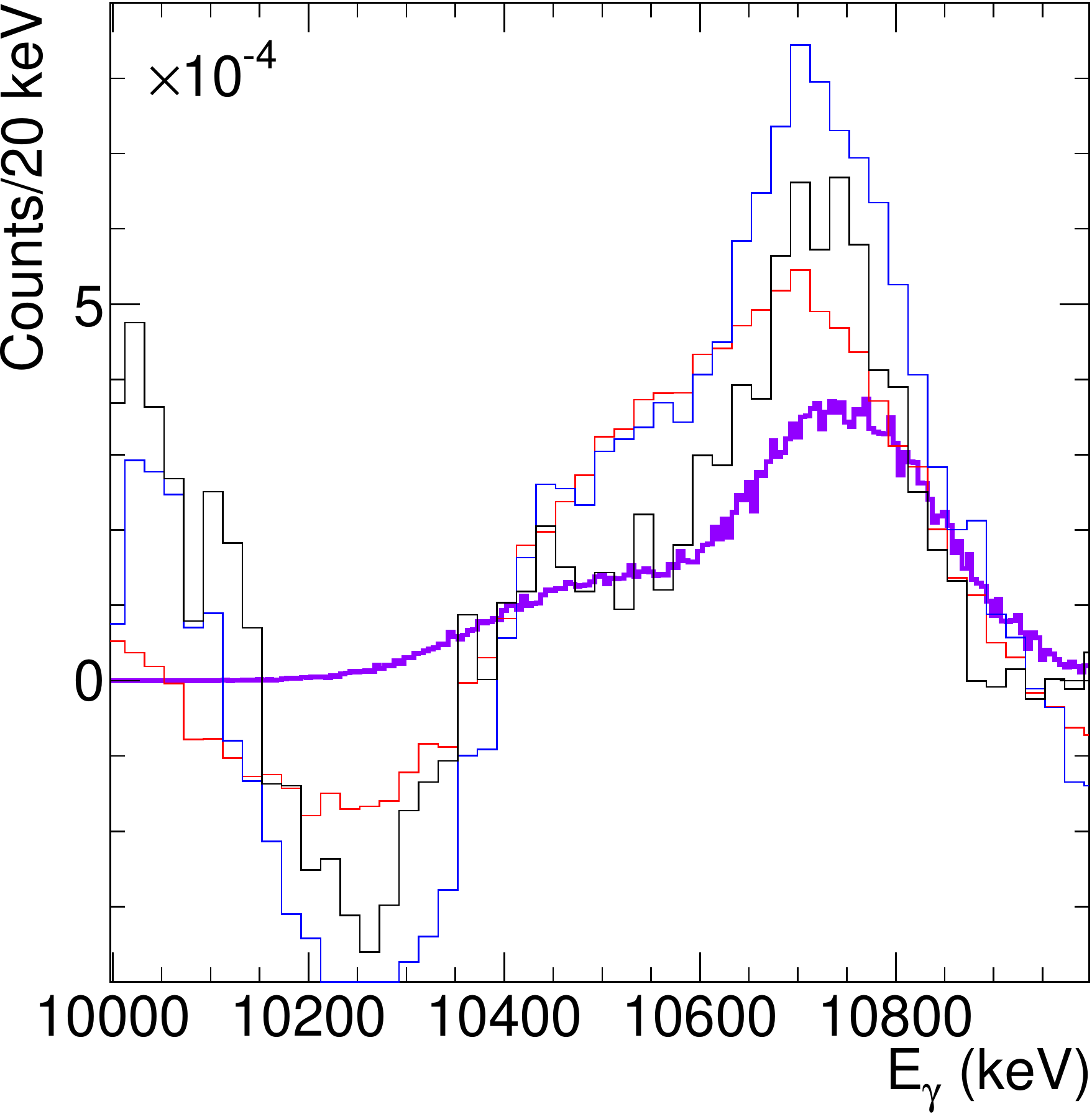}
 \caption{Performance of the unfolding algorithms with a subset of the data corresponding to a reduction of a factor of $10^{-2}$ (left), $10^{-3}$ (middle), and $10^{-4}$ (right) in statistics. The black line correspond to the iterative method, red line correspond to Matrix L1, and blue line correspond to Matrix L2. The purple spectrum shows HPGe data obtained in similar experimental conditions, refolded with the \labr\ full-energy peak response and renormalized to the same intensity.\label{fig:stats}}
\end{figure*}

To investigate the effects of statistical fluctuations on the performance of the different algorithms we have repeated the unfolding processes for three different sub-sets of the data. These three sub-sets were selected to contain a factor of $10^{-2}$, $10^{-3}$, and $10^{-4}$ of the events. This correspond to a peak height for the full-energy peak of 100 counts, ten counts, and one count with bin sizes of 5~keV, respectively. The results of these procedures are shown in Figure~\ref{fig:stats}, where the spectra have been re-binned for clarity. For the first case the performance of the three methods is similar to the case with full statistics. In all three cases the iterative method generally follow the trend of the re-folded \ac{HPGe} spectrum. However, with a data reduction of $10^{-3}$ large-scale oscillations start to appear for the second-derivative regularization and the iterative method. All three methods overestimate the size of the peak in the extreme case with a $10^{-4}$ reduction of the data. As a general trend, the Thikhonov-Phillips regularization appears as the most stable method in the low-statistics limit.

\section{Conclusions}

We have performed a qualitative and quantitative comparison between three different unfolding schemes for $\gamma$-ray spectra from \labr\ detectors in the high-energy regime. The results show that, for $\gamma$-rays around 10~MeV the iterative approach gives more reliable reconstruction of hidden fine-structures in the emission spectrum as well as smaller correlated fluctuations of the data related to the unfolding process. The results of the unfolding, with emphasis on the identification of hidden structures, is furthermore parameter free in the iterative approach and does, thus, not depend on the somewhat arbitrary choice of a regularization parameter. In addition, this approach has the advantage of being computationally relatively fast, making it suitable for fast evaluation of semi on-line data for physics experiments and diagnostics of a $\gamma$-ray beam.

Finally we would like to add a  general note  of caution that the choices of type of regularization, as well as the strength of the regularization, depends on the type of data and analysis that is performed. All three cases presented here will introduce a bias to the unfolded data, especially if over-regularized, or if the iterative procedure is stopped too early. This report has discussed the unfolding strategies in terms of sparse high-energy spectra from \labr\ detectors. For practical implementation in other setups, for example providing spectra with very narrow peaks and sharp structures or smooth low-resolution spectra with high $\gamma$-ray densities, it is important that the chosen method is evaluated for the intended application.

\acknowledgments

The authors would like to acknowledge the support from the Extreme Light Infrastructure Nuclear Physics (ELI-NP) Phase II, a project co-financed by the Romanian Government and the European Union through the European Regional Development Fund - the Competitiveness Operational Programme (1/07.07.2016, COP, ID 1334). We acknowledge A.~Imreh from the Technical Division at ELI-NP for the CAD drawings of the detector system used for \tfig\ref{fig:setup}.

%


\providecommand{\href}[2]{#2}\begingroup\raggedright\endgroup

\acrodef{AdaGrad}{adaptive gradient algorithm}
\acrodef{AGATA}{Advanced GAmma Tracking Array}
\acrodef{ANN}{artificial neural network}
\acrodef{API}{application programming interface}
\acrodef{BRIKEN}{Beta-delayed neutrons at RIKEN}
\acrodef{CFD}{constant-fraction discriminator}
\acrodef{CLHEP}{Class Library for High Energy Physics}
\acrodef{DAQ}{data acquisition}
\acrodef{ELI-BIC}{ELI Bragg ionization chamber}
\acrodef{ELI-NP}{Extreme Light Infrastructure -- Nuclear Physics}
\acrodef{ELI}{Extreme Light Infrastructure}
\acrodef{ELIADE}{ELI Array of DEtectors}
\acrodef{ELIGANT}{ELI Gamma Above Neutron Threshold}
\acrodef{ELIGANT-GG}{ELIGANT Gamma Gamma}
\acrodef{ELIGANT-GN}{ELIGANT Gamma Neutron}
\acrodef{ELIGANT-TN}{ELIGANT Thermal Neutron}
\acrodef{ELISSA}{ELI Silicon Strip Array}
\acrodef{EOS}{equation of state}
\acrodef{FADC}{flash ADC}
\acrodef{FASTER}{Fast Acquisition System for nuclEar Research}
\acrodef{FATIMA}{FAst TIming Array}
\acrodef{FFT}{fast Fourier transform}
\acrodef{FWHM}{full-width at half-maximum}
\acrodef{GBS}{gamma-beam system}
\acrodef{GDR}{giant dipole resonance}
\acrodef{GROOT}{\geant\ and ROOT Object-Oriented Toolkit}
\acrodef{GSI}{Gesellschaft f\"{u}r Schwerionenforschung}
\acrodef{HPGe}{high-purity germanium}
\acrodef{HPLS}{high-power laser system}
\acrodef{HV}{high voltage}
\acrodef{IAEA}{International Atomic Energy Agency}
\acrodef{IFIN-HH}{Horia Hulubei National Institute of Physics and Nuclear Engineering}
\acrodef{JINR}{Joint Institute for Nuclear Research}
\acrodef{kBq}{kilo bequerel}
\acrodef{keV}{kilo electron-volt}
\acrodef{MCNP}{Monte Carlo N-Particle Transport Code}
\acrodef{MIDAS}{Multi Instance Data Acquisition System}
\acrodef{ORNL}{Oak Ridge National Laboratory}
\acrodef{PCI}{Peripheral Component Interconnect}
\acrodef{PDR}{pygmy dipole resonance}
\acrodef{PMT}{photomultiplier tube}
\acrodef{PRD}{prompt-response distribution}
\acrodef{QPM}{quasiparticle-phonon model}
\acrodef{QED}{quantum electrodynamics}
\acrodef{RA1}{Research Activity 1 - ``High-Power Laser System''}
\acrodef{RA2}{Research Activity 2 - ``High-Brilliance Gamma Beam''}
\acrodef{RA3}{Research Activity 3 - ``Nuclear Physics with High-Power Lasers''}
\acrodef{RA4}{Research Activity 4 - ``Nuclear Physics and Applications with high-brilliance gamma-beams''}
\acrodef{RA5}{Research Activity 5 - ``Fundamental Physics with combined laser and gamma beams''}
\acrodef{ReLU}{rectified linear activation unit}
\acrodef{RMS}{root-mean-square}
\acrodef{ROSPHERE}{ROmanian array for SPectroscopy in HEavy ion REactions}
\acrodef{TDR}{technical design-report}
\acrodef{TOF}{time-of-flight}
\acrodef{UPC}{Universitat Polit\`{e}cnica de Catalunya}
\acrodef{VME}{Versa Module Europa}

\end{document}